\documentstyle[12pt,aasms4]{article}
\begin{document}
\def\Msun {M_{\odot}\ }
\def\msun{{\rm\,M_\odot}}
\def\lsun{{\rm\,L_\odot}}
\def\rsun{{\rm\,R_\odot}}
\title{A WFPC2 Study of the Resolved Stellar Population of the
Pegasus Dwarf Irregular Galaxy (DDO~216)\altaffilmark{1}}
\vskip0.5cm
\author{J. S. Gallagher}
\affil{University of Wisconsin, Department of Astronomy, 475 N. Charter St., 
Madison, WI 53706-1582\altaffilmark{2}}
\author{E. Tolstoy}
\affil{ST-ECF, Karl-Schwarzschild Str. 2, D-85748 Garching b. M\"unchen, 
Germany}
\author{Robbie C. Dohm-Palmer, E. D. Skillman}
\affil{University of Minnesota, Department of Astronomy, Minneapolis, MN 55455}
\author{A. A. Cole and J. G. Hoessel}
\affil{University of Wisconsin, Department of Astronomy, 475 N. Charter St.,
Madison, WI 53706-1582}
\author{A. Saha}
\affil{Space Telescope Science Institute, 3700 San Martin Dr., Baltimore, 
MD 21218}
\author{M. Mateo}
\affil{University of Michigan, Department of Astronomy, 821 Dennison Building, 
Ann Arbor, MI 48109-1090}
\altaffiltext{1}{Based on observations with the NASA/ESA {\it Hubble Space 
Telescope} obtained at the Space Telescope Science Institute, which is 
operated by AURA, Inc. under NASA contract NAS 5-26555}
\altaffiltext{2}{email: jsg@tiger.astro.wisc.edu}
\tighten
\begin{abstract}
The stellar population of the Pegasus dwarf irregular galaxy is 
investigated in images taken in the F439W (B), F555W (V), and F814W (I) 
bands with the Wide Field Planetary Camera 2 (WFPC2) on the {\it 
Hubble Space Telescope}. With WFPC2 the Pegasus dwarf is highly 
resolved into individual stars to limiting magnitudes of about 25.5 
in B and V and 25 in I.  These and ground-based data are combined 
to produce color-magnitude diagrams which show the complex nature of 
the stellar population in this small galaxy. A young (age $<$ 0.5~ Gyr) main 
sequence stellar component is present and clustered in two centrally-located 
clumps, while older stars form a more extended disk or halo. The 
colors of the main sequence require a relatively large extinction 
of A$_V =$ 0.47 mag. The mean color of the well-populated red giant 
branch is relatively blue, consistent with a moderate 
metallicity young, or older, metal-poor stellar population. 
The red giant branch also has significant width in color, implying a 
range of stellar ages and/or metallicities.
A small number of extended 
asymptotic giant branch stars are found beyond the red giant 
branch tip. Near the faint limits of our data 
is a populous red clump superimposed on the red giant branch.
Efforts to fit self-consistent stellar population models based on 
the Geneva stellar evolution tracks yield a revised distance of 
760 kpc. Quantitative fits to the stellar population are explored as 
a means to constrain the star formation history. 
The numbers of main sequence and core helium-burning blue loop 
stars require that the star formation rate was higher in the 
recent past, by a factor of 3-4 about 1~Gyr ago. 
Unique results cannot 
be obtained for the star formation history over longer time 
baselines without better information on stellar metallicities and 
deeper photometry. 
The youngest model consistent with the data contains stars with 
constant metallicity of Z$=$0.001 which mainly formed 
2-4 Gyr ago. If stellar metallicity declines with 
increasing stellar age, then older ages are allowed of up 
to $\approx$8~Gyr. However, even at its peak of 
star forming activity, the intermediate-age dominated model for 
the Pegasus dwarf most likely remained relatively dim with 
M$_V \approx -$14.

\end{abstract}

\section{Introduction}

The Pegasus dwarf irregular galaxy  (DIG) was first identified by 
A. Wilson (see Holmberg 1958) 
and its status as a nearby dwarf was confirmed only
with the detection of HI by Fisher \& Tully (1975).  Despite its small
size, Pegasus supports ongoing star formation as evidenced by
the presence of luminous blue stars (Hoessel \& Mould 1982, 
Christian \& Tully 1983,
Sandage 1986) and small HII regions (Hunter,
Hawley, \& Gallagher 1993). Pegasus
is fairly typical of the least luminous DIGs detected in
the Local Group and a few other nearby galaxy groups (cf. Karachentseva
et al. 1987;
Miller 1996). 

Pegasus presents an interesting structural combination of a 
dwarf irregular system, with a chaotic appearance due to star formation, 
and a relatively symmetric dwarf elliptical (dE) or dwarf spheroidal 
(dSph) galaxy, typically without young stars.  Pegasus 
has a smooth outer envelope with elliptical isophotes and a 
brighter core that contains at least two OB associations (Ivanov 1996).  
It also has a relatively low amount of HI, 
as indicated by its moderate (for a DIG) ratio of M$_{HI}/$L$_V =$0.4 
(Hoffman et al. 1996). 
Pegasus may therefore be a nearby example of a transition object between 
a dwarf galaxy dominated by current
star-formation and one dominated by past star formation.  

While the evolution of very low mass galaxies is 
still unclear, one possibility is that brief  
epochs of very active star formation produce a significant fraction 
of the stellar mass over a 
few dynamical time scales, or $<$1~Gyr (e.g. Babul \& Ferguson 1996). 
Such events could 
disrupt or eject a small galaxy's ISM,  
thereby inhibiting further star formation for a time (e.g. 
Dekel \& Silk 1986, Sandage \& Fomalant 1993, Marlowe et al. 1995). 
Under these conditions much of the metal-rich supernova ejecta will 
also be lost and so the remaining stellar populations are expected 
to be relatively metal-poor and may show little metal enrichment 
over time.  It is interesting to see 
if any indications exist for epochs of enhanced 
star formation activity in the recent history of 
the Pegasus dwarf, which might have led to its transitional 
morphological structure. 

This paper presents the first optical study of the stellar population of the
Pegasus dwarf based on observations obtained with the Wide
Field Planetary Camera 2 (WFPC2) on the {\it Hubble Space Telescope}
({\it HST}). Supplementary ground-based data were obtained with the 
WIYN\footnote{The WIYN Observatory is a joint facility of the 
University of Wisconsin-Madison, Indiana University, Yale University, 
and the National Optical Astronomy Observatories.}
3.5-m telescope. The Pegasus dwarf is highly resolved into individual 
stars by WFPC2.  These data sets allowed us to obtain high quality  
color-magnitude diagrams (CMDs) for the Pegasus dwarf, which we analyze   
to derive its recent star formation history (SFH). 

The next section reviews the global properties of Pegasus
and \S3 describes our new observations.
Our analysis of the color-magnitude diagrams is covered
in subsequent sections, the derived SFH is presented in \S7, 
and the results are summarized in \S8.

\section{Global Properties of the Pegasus Dwarf Irregular Galaxy}

\subsection{Stellar Populations}

Table 1 summarizes several key observables for Pegasus. The 
B$-$V color of Pegasus is sufficiently blue that some 
ongoing star formation is
required. This is consistent with the CMD from
an early CCD study by Hoessel \& Mould (1982).
They noted a modest young stellar population component
and suggested that ``...recent star formation in Pegasus has been very
subdued.''

This point was reinforced by the discovery (Hunter, Hawley, \&
Gallagher 1993) that Pegasus contains small, faint HII regions
and some diffuse ionized gas\footnote{A narrow-band 
H$\alpha$ image obtained with the WIYN Telescope in 
August 1997 under good seeing 
conditions by Wisconsin graduate students 
A. Cole, J. C. Howk, \& N. Homeier (private communication)  
suggest the second of the original 
HII regions is a background galaxy. These data also show that Pegasus 
contains several previously undetected faint, compact H$\alpha$ sources 
which could be HII regions.} (see
also Aparicio \& Gallart 1995, 
Skillman, Bomans, \& Kobulnicky 1997). Based on 
the Hunter et al. observed
L($H\alpha$)$=$1 $\times$ 10$^{36}$ (D/1 Mpc)$^2$ erg~s$^{-1}$, 
we estimate that only a few late type O stars have been born in the
past few Myr. This leads to a rough estimate of the
current star formation rate (SFR$_0$)
for a Salpeter initial mass function
of  $\rm \dot M_* \approx 3 \times 10^{-4} \Msun~yr^{-1}$ or 
SFR$_0 \sim$500~$\Msun$~kpc$^{-2}$~Myr$^{-1}$ averaged over the 
central region of the Pegasus dwarf. 
For a stellar mass of
$\rm M_* = 7 \times 10^6$(M/L$_V$) $\Msun$, 
the Roberts time to form the existing
stars at the current SFR would be $>$20 Gyr for the expected 
M/L$_V \geq$ 1. The low L($H\alpha$), lack of luminous blue stars, 
and B$-$V color 
indicate that the SFR$_0$ is 
lower than the lifetime mean SFR for this galaxy. 

More recently Aparicio \& Gallart (1995) and 
Aparicio, Gallart \& Bertelli (1997a) made a 
ground-based study of Pegasus based on VRI photometry of
resolved stars. They confirm the presence of young, evolved red stars. 
These are seen against an extensive background population of older stars
on the red giant branch (RGB),  which they suggest have ages of up to about 10
Gyr, and a range of possible SFHs 
are derived.  However, CMDs obtained 
from ground-based observations are necessarily 
limited in precision and depth due to image crowding. Stellar 
images observed with WFPC2 typically 25 times smaller in angular area 
than those measured from the ground even in excellent seeing.  WFPC2 therefore 
yields more precise CMDs of the crowded 
stellar populations of galaxies,  which are 
essential for constraining SFHs.

\subsection{HI and Kinematics}

Lo, Sargent, \& Young (1993) mapped Pegasus with the Very Large Array in
the HI 21-cm line. Their maps show that the HI is concentrated within the
core of the optical body of the galaxy, where several distinct clumps
are present.  The mean HI column density is $N(HI) =$10$^{21}$~cm$^{-2}$
and the peak values are at least twice the mean.
The peaks roughly coincide with the region occupied by blue stars.
A comparison with single-dish maps by Hoffman et al. (1996) shows that the
Lo et al. VLA HI maps contain 
most of the HI flux; Pegasus does not appear to 
have an extended, massive  reservoir of interstellar HI gas in an outer 
disk or halo.

\subsection{HII Regions and ISM Abundances}

Spectroscopy of the brightest HII region in Pegasus
has been obtained by Skillman,
Bomans, \& Kobulnicky (1997). The low excitation
spectrum shows strong [O II] $\lambda$3727 emission and [O III]
$\lambda$5007 emission is not detected. 
Using these data, Skillman et al. derive an
abundance of 12$+$log(O/H)$=$7.9$\pm$0.1--0.2 (Z=0.002 or 10\% solar); 
an oxygen abundance which
is close to that of the Small Magellanic Cloud. Thus Pegasus appears to
be surprisingly 
metal-rich relative to its blue luminosity. From the emission line
Balmer decrement Skillman et al. derive an extinction of 
E(B$-$V)$=$ 0.2 $\pm 0.1$, which is much larger than the 
E(B$-$V)$=$0.03 predicted by 
the standard Galactic extinction model of Burstein \& Heiles (1984).
Reddening plays an important role in interpreting the CMDs of the 
Pegasus dwarf galaxy, and we discuss this issue in more detail in \S3.3.

\section{Observations}

\subsection{Observations with {\it HST} Wide Field Planetary Camera 2.}

Our WFPC2 data consist of three exposures of 600 s each in 
the F555W (V) and F814W (I) filters,  and a set of 2$\times$900~s and 
2$\times$ 1100~s  exposures in the F439W (B) filter.
Each image is offset by a few pixels plus a fractional
pixel offset, or
``dithered''  (Leitherer 1995, Fruchter \& Hook 1996) 
with respect to each other in an attempt to compensate
for the under-sampled point spread function of WFPC2. The
images in each filter were first registered 
to the nearest integer pixel, and then 
cosmic-ray cleaned and combined using techniques 
described by Saha et al. (1996).  The nominal pointing was to 
$\alpha_{2000}=$23:28:33.0, $\delta_{2000}=+$14:44:6.0 at the 
WFALL-FIX aperture which is located approximately 8 arcsec along 
the diagonal running from the center of the WFPC2 field to the 
corner of the WF3 camera.
The resulting 
WFPC2 V image of the heart of Pegasus is shown in 
Figure 1 with a close-up view in Figure 2. 

Photometry was carried out on the combined images using a version of
DoPHOT (Schechter, Mateo, \& Saha 1993) altered to  
take account special circumstances presented by under-sampled 
WFPC2 images,  which also contain variable point spread functions
(Saha et al. 1996).
The photometry was calibrated and converted to the ``standard''
BVI system using the precepts laid out in 
Holtzmann et al. (1995). Independent analysis of the images 
indicate zero point uncertainties 
of $\pm$0.05~mag in V and $< \pm$0.03~mag in V$-$I. We 
therefore adopted these values as our minimum photometric errors.   
The B-band data show larger scatter, reflecting their 
their lower signal-to-noise ratios. 
The resulting color-magnitude diagrams are presented in 
Figures~3 and 4. for the individual WFPC2 CCDs and for the combined 
data in Figure 5.  Figure 6 displays the internal
photometric error distributions.  The properties of these 
CMDs are described in \S4.

\subsection{Ground-Based Observations with the WIYN Telescope}

CCD images in the V and I filters were obtained in September 1996 with the 
WIYN 3.5-m telescope.  The field of view of the 2048$\times$2048 pixel 
CCD is 6.7 arcmin at a scale of 0.2 arcsec per pixel.  The exposure times were 
600 sec in I and 500 sec in V.  Conditions were clear around Pegasus 
but deteriorating elsewhere in the sky during our observations; we cannot be 
sure of our photometric zero points due to the possible presence of clouds. 
The full width at half-maximum image sizes were $\approx$0.6 arcsec.

These images were processed in a standard way with 
IRAF\footnote{IRAF is distributed by the National Optical Astronomy 
Observatories, which are operated by the Association of Universities 
for Research in Astronomy, Inc., under cooperative agreement with the 
National Science Foundation.}. 
Transformations to 
standard magnitudes were approximated by matching our WFPC2 results. 
The WIYN telescope V-band image is shown in Figure 7.   
Results of our DoPHOT photometry 
of the WIYN images are displayed in Figure 8.  
The main sequence (MS) is more scattered and the RGB is broader in 
color in Figure 8 
than in the HST CMD in Figures~4, consistent with the expected 
effects from crowded ground-based images.
There is also a larger population of bright red (I$<$21 \& V$-$I$>$1.5)
stars, on the extended-asymptotic giant branch (AGB), 
which the WIYN data show extend beyond the radius in the Pegasus dwarf 
covered by our WFPC2 images. 

\section{The Color Magnitude Diagrams}

\subsection{Description of Major Features}

A sparsely-populated 
 ``blue plume'' of well-resolved stars is present in our CMDs and is  
best defined in the WFPC2 V, B$-$V CMD in Figure~3. It consists 
of a MS with adjacent core helium burning (HeB)
evolved  intermediate mass stars which form the `blue loop'.

The straight, nearly vertical structure of the 
MS indicates that
the hotter young stars in the Pegasus dwarf 
have spectral classes of A or earlier; at later
spectral types the MS changes to a flatter slope
where color varies more rapidly with luminosity (c.f. the
Large Magellanic Cloud field observed by Gallagher et al.
1996).  We can obtain an idea of the 
recent SFH through comparisons with other 
nearby star-forming 
galaxies observed with WFPC2, 
such as the Magellanic Clouds (Gallagher et al. 1996, Holtzman 
et al. 1997),  Sextans A 
(Dohm-Palmer {\it et al.} 1997a,b), and Leo~A (Tolstoy et al. 1998).
These confirm that even by the standards of low 
surface brightness 
dwarf irregular galaxies, Pegasus has a relatively small population 
of young stars.

The dominant feature of the I,V$-$I CMD (Figure~5) is the populous
`red plume', consisting of the RGB and
a dense concentration of core helium-burning stars 
in the red clump (RC) overlying the lower RGB. 
In addition
we see members of an extended-AGB population, and this is
confirmed in the larger field WIYN data (see also Aparicio, Gallart, 
\& Bertelli 1997a).  A few 
red supergiant (RSG) stars are likely present with colors 
of V$-$I$\approx$1 and I$<$21. In the blue the RGB is spread 
in color and not well-measured at the depth reached by our 
WFPC2 data. This part of our CMD resembles the WFPC2 
observations of the Local Group dE galaxy 
NGC~147 (Han et al. 1997).
The RGB and the RC are prominent in both galaxies  
(see Figure~9), which is surprising given their  
very different morphological 
classes.   The obvious differences between the CMDs of the 
two galaxies are the horizontal branch (HB), which is
seen in NGC~147 but not in the shallower 
Pegasus WFPC2 images; the greater color width 
of the RGB in NGC~147; the sharp RGB tip in NGC~147; and the 
presence of young stars in Pegasus.
The older stellar population appears to be more complex in terms 
of its range of ages and metallicities in NGC~147 than in Pegasus, 
but star formation has persisted for a longer time in the smaller 
Pegasus system.

\subsection{Spatial Distributions}

Spatial density distributions of stars also provide a means to explore 
the SFH of galaxies. For example, it is standard practice to use 
OB stars as markers of recent star formation, and this 
philosophy can be readily extended to less luminous main 
sequence stars to investigate stellar age-spatial distribution 
correlations in galaxies. Additional insights into the SFH  
are sometimes supplied by star clusters, whose ages can 
be reasonably well-determined from colors when they are  
less than a few Gyr old (e.g. Hodge 1980).  

Unfortunately in the region of the Pegasus dwarf 
covered by WFPC2 we find only one half of a moderately dense  
star cluster. Possible star clusters were noted by Hoessel \& Mould 
(1982).  The WFPC2 images show that the Pegasus dwarf is in front 
of a moderately rich group of galaxies, and some of 
these background galaxies could have been mistaken for star clusters 
in ground-based images. 
However, the central star cluster is partially in our PC image, and 
has a diameter of $\approx$40~pc with  
a moderate stellar density enhancement over the surrounding field.
We attempted to make a CMD for this cluster but it is of 
poor quality due to crowding and otherwise indistinguishable 
from the surrounding field. 
The cluster appears to have a strong RC and RGB with no 
definite MS stars; thus it is 
probably an intermediate age object, older than 
about 2~Gyr. A deeper observation including the entire cluster could yield a 
more accurate cluster CMD and allow a better age estimate.

While the connections between SFRs and star cluster formation rates 
remain obscure, very dense, luminous `super star clusters' 
are often produced in small galaxies during  
starbursts (Meurer et al. 1992, O'Connell et al. 1994). 
These types of clusters would be denser and richer than 
the Pegasus central star cluster.  The absence 
of candidate super star clusters in Pegasus is consistent with 
models where no starburst occurred 
during at least the past few Gyr, this time 
scale being set by the likely minimum dynamical lifetimes for dense star 
clusters to dissolve (Goodwin 1997). 

The general pattern in all nearby galaxies 
is for younger stars to clump into associations, although 
examples of young, massive stars in the field also exist 
(e.g. Massey 1997). Stellar associations drift apart with velocities 
of a few km~s$^{-1}$, and former 
members will diffuse within a galactic disk; the 
mixing times of stars within the disks of galaxies 
are typically $\leq$ 1 Gyr (see \S5.1 in Gallagher et al. 1996). 
In addition, the region of a galaxy that is accessible to stars 
born in a disk 
may be bounded by the energy and angular momentum of their 
parent interstellar gas clouds. Thus stars formed from interstellar clouds 
near the center of a galaxy are unlikely to have sufficient 
energy or angular momentum to travel very far out in a stellar disk. 

Stellar surface density distributions derived from our 
WFPC2 images of the Pegasus 
dwarf are shown in Figure~10.  The stellar populations 
in the Pegasus dwarf show the expected trend 
for most of the younger stars to be spatially clumped, and the older stars 
to be more smoothly distributed. Figure~10a 
illustrates the surface densities of young stars located in 
the blue plume from the V, B$-$V CMD in Figure~4a. 
These stars are in two major concentrations which 
resemble diffuse OB associations (Ivanov 1996), and are located near 
the center of the optical galaxy. As shown in Figure~10b, 
the RGB and RC stars are more symmetrically distributed
about the central star-forming zone.  

Figure 10c presents the density of extended-AGB stars derived from the WIYN 
CCD images. We find good agreement between the surface densities of 
extended-AGB stars measured in the WFPC2 images and from WIYN; we 
therefore prefer the wide field of view WIYN images for this plot.
The densest concentration of extended AGB stars is found along the 
major axis, following approximately the same pattern as is seen 
in all of the other stars. Exterior to this zone the distribution 
of extended AGB stars is lumpy; we may therefore be seeing remnants 
of older star forming regions or the modulation of apparent stellar 
densities by dust within the Pegasus dwarf. The outermost regions 
appear as discrete clumps due to statistical fluctuations.

\section{Ingredients for Modeling the CMD}

\subsection{Stellar Evolution Models}

Stellar evolution tracks give the luminosity, effective 
temperature, and surface gravity for single stars of a given mass 
and chemical composition as a function of age.  
The tracks depend on basic astrophysical parameters, 
such as the initial mass, metallicity level and distribution.
The properties of real stellar populations also depend on a 
variety of properties not explicitly included in most current 
stellar evolution models, such as stellar rotation rates
and the presence of close binary companions.  The results from numerical
calculations are also sensitive to the treatment of convection
in terms of the ratio of pressure scale height to convection
scale and to the effect of convective core overshoot.

In this paper we primarily use two sets of stellar evolution
tracks at Z=0.004 and Z=0.001 computed by the Geneva group 
(Charbonnel et al. 1993, Schaller {\it et al.} 1992), which we 
compare with the tracks given by the Padua group stellar evolution 
models (Fagotto et al. 1994). 
In order to most accurately compare the two
sets of model tracks, it is necessary to work from a uniform
set of equivalent evolutionary points (EEP).  We chose to use the
EEP defined by Schaller et al. (1992).  These
points are defined differently during the various phases of 
stellar evolution.  For example, on the main-sequence, the 
EEP represent a sequence of decreasing central hydrogen abundance.
The EEP must be carefully chosen so as to properly sample all
of the color-magnitude space spanned by the tracks. This ensures
the ability to correctly model, e.g, the ``bump'' in the RGB 
luminosity function, or the main-sequence ``hook'' exhibited by
stars with convective cores.  

In converting the Padua tracks
to the Geneva EEP, we carefully interpolated the tabulated values
of {\sl log}~L, {\sl log}~T$_{eff}$, and age at the given points
to the appropriate values at the EEP.  Along each 
Padua track we identified the 51 Geneva EEP 
and interpolated new tracks between these points using a cubic spline. 
The accuracy of this process 
was checked by over-plotting the original and converted
tracks; the differences were found to be negligible.
This allows us to properly 
interpolate when producing stellar population models, as discussed by
Tolstoy (1996).  While we do not explicitly include binary stars in our 
models, we do allow for a population of outlying stars to be present,  
which effectively account for binary star blends in our data.

The two sets of stellar evolution
models differ in several details, as can be seen in Figure~11 where
we plot the 1.5$\Msun$ and 2$\Msun$
stellar evolution tracks from Padova (solid lines) and
Geneva (broken lines). The RGB tip luminosity
differs by more
than 0.5 magnitude, and on the upper part of the 
RGB, the Padova tracks are frequently
several tenths of a magnitude bluer than the Geneva tracks. 
These differences are probably due to the treatment of convective 
overshoot or the exact definition of the point where the helium-flash 
occurs and the RGB models are terminated.
The agreement
between the tracks is better along the MS and
at the base of the RGB. {\it Therefore, if
we used Padova rather than Geneva stellar evolution tracks we would find 
a slightly larger distance and a younger (or more metal poor, or both)
stellar population.} 

The {\it numbers} of stars predicted
to exist at any location off of the zero-age MS 
for the same SFH also differs for
Padova and Geneva stellar evolution models. This occurs because 
at a given initial mass, the Padova stellar models are generally 
more luminous and also have longer lifetimes. The initial mass 
function requires that the numbers of stars born per unit stellar 
mass will increase as stellar mass decreases. For a given 
SFR the Padova tracks will predict more stars at 
each luminosity (after the zero age MS) than the 
Geneva tracks. {\it As a result, the SFR derived from an observed 
CMD will be somewhat lower for Padova models than for the Geneva models.}

Because most galaxies have complex SFHs, the many
interconnected effects that determine the properties of a 
galactic CMD come into play.   
The history of a galaxy can therefore be 
effectively modeled using numerical 
(e.g. Monte Carlo) simulations, where a composite 
stellar population is randomly 
extracted from theoretical stellar evolution tracks using an assumed 
Initial Mass Function (IMF) and SFH. 
We follow the approach described in detail by Tolstoy~(1996), and 
made a number of simulated CMDs for a variety of assumed SFH. 
Each simulation then depends upon {\it all}
the main parameters which  determine an observed CMD, unlike the standard
isochrone fitting methods  previously used.
Monte Carlo simulations were first used to simulate galactic CMDs by 
Tosi, Greggio and collaborators (e.g. Tosi et al.~1991), and these
ideas were further developed by Tolstoy \& Saha (1996) and Tolstoy (1996)
to allow statistical comparisons between many different CMD 
models and the multiple color photometric data sets.

Using the techniques of Tolstoy \& Saha (1996) we created 
stellar population models for a variety of potentially 
feasible SFH 
using Geneva stellar evolution tracks. These are comparatively 
well-defined for the recent history of the Pegasus dwarf, 
but become non-unique at ages beyond about 2 Gyr. 
We then determined the most
likely SFH to match our observed CMD for the Pegasus dwarf.
Our method allows for limitations in the data, 
such as the increasing scatter and rising incompleteness at fainter 
magnitudes, which preclude finding 
a unique solution for the SFH of Pegasus based on our current data sets.

\subsection{Matching the Observational Errors}

The two main observational effects that we  must incorporate into 
our model CMDs before we can accurately
compare them to observed CMDs are
observational errors (plotted in Figure~6), and the incompleteness,
or the number of stars we miss at given observed magnitude range
because of crowding  or noise (cf. Tolstoy~1996). 
Because our HST images are effectively
uncrowded the incompleteness is very well correlated with the
observational errors. This issue has been studied for
Sextans~A in detail by Dohm-Palmer et al. (1997a), and
we use these results in our models.

\subsection{Extinction}

The standard Galactic extinction to the Pegasus dwarf derived by Burstein \& 
Heiles (1984) is small, E(B$-$V)$=$0.03. However, 
our observations show that with this extinction correction, 
the Pegasus dwarf's MS is too 
red in B$-$V and V$-$I; we could find no models with low reddening 
that would simultaneously fit in both colors. We made a model  
B$-$V versus V$-$I two color diagram, and empirically derived a 
possible range of 
reddenings (and distances, with which it is weakly correlated).
Feasible reddening values were identified by 
requiring the model and observed
MS to agree to within $\pm$ 0.05 magnitudes in both B$-$V and
V$-$I colors.  This result is almost independent of our choice of 
abundance or recent SFH, since it depends only on the almost invariant 
optical colors of the upper main sequence. 
We also required fits at 0.2 magnitude level to the
RGB tip and to the RC, while avoiding producing
CMD features which are {\it not} seen, such as an extended blue
loop. We thus find E(B$-$V)$=$0.15$\pm0.05$ from model fits to 
our WFPC2 BVI CMDs. 

To check the consistency of this reddening derived from the CMDs we
redetermined the mean Galactic extinction {\it towards} Pegasus by
converting the Galactic HI column density to an extinction. For this test 
we took $N(HI) =$3.7$\times$10$^{20}$~cm$^{-2}$ derived 
from an HI spectrum recently taken  by Kalbera (1997) with the 
Effelsberg  
100-m radio telescope which has a 9~arcmin beam size. 
Following Diplas \& Savage (1994), we find that
E(B$-$V)$=$0.08 is predicted for a standard Galactic dust-to-gas
ratio. The {\it internal} extinction
within Pegasus should be low (Fitzpatrick~1995) unless the mean $N(HI)$
is severely underestimated. Using the Lo et al. average $N(HI)$ 
within Pegasus, the internal extinction for an SMC dust-to-gas ratio would be
E(B$-$V)$\leq$0.03. from the HI data.  We therefore estimate a total 
extinction of about E(B$-$V)$ \approx$ 0.1 from the measured $N(HI)$. 

Additional information comes from the IRAS 
60$\mu$m and 100$\mu$m FRESCO maps of the region around Pegasus.
If there is a large amount of FIR (100$\mu$m excess) cirrus in front
of Pegasus, the HI observations could under-estimate the Galactic
extinction. The 100$\mu$m IRAS
map is shown in Figure~12, where we see a small FIR excess
at the position of Pegasus. The 100$\mu$m/60$\mu$m ratio
is consistent with the more general infrared cirrus in this field; 
this is not a dust-free direction. 
A quantitative estimate of the extinction from the IRAS data 
can be made following the approach of Laureijs, Helou, \& Clark (1994). 
For a peak 100$\mu$ brightness of 0.4 Jy steradian$^{-1}$ 
averaged over the $\approx$6~arcmin effective resolution of 
the IRAS map, the predicted 
extinction is A$_B =$0.03 mag. However, we might expect this 
to be an underestimate if the dust is far from the heating sources 
in the Galactic disk, 
and therefore cold and radiating inefficiently in the IRAS 100$\mu$ band.
The high 60$\mu /$100$\mu$ ratio of the dust towards Pegasus supports  
this possibility.

The estimators for foreground Galactic extinction 
toward the Pegasus dwarf do not yield consistent 
values, and furthermore are averages over larger angular 
scales then the size of the WFPC2 field of view. 
Since we are working from CMDs, our data require a larger 
extinction than the Burstein and Heiles value. We therefore adopt 
E(B$-$V)$=$0.15 for the remainder of this paper and apply this extinction 
to all of our data: A$_B=$0.62, A$_V=$0.47, and A$_I=0.28$

\subsection{Distance}
 
The distance to the Pegasus DIG has not been
independently established from well-calibrated standard candles, 
such as Cepheid or RR Lyrae variable stars.  
Hoessel et al. (1993) reported the detection of what appeared to be
Cepheid variable stars 
in Pegasus from a series of CCD images taken in the Thuan-Gunn {\it r}
band.
Their distance modulus was (m-M)$_0 =$ 26.2 corresponding to
D$=$1.7~Mpc.  Aparicio (1994) showed 
the Hoessel et al. candidate Cepheids to have 
the  red colors of stars associated with the RGB. 
Photometry from our WIYN images agrees with his conclusion;
the ``Cepheids'' 
are some other class of red variable star and the Cepheid distance 
is not valid.  Aparicio derived a revised distance modulus of
(m-M)$_0 =$24.9 $\pm$ 0.1, or D$=$960~kpc, from the location of the RGB tip 
following the calibration of Lee, Freedman, \& Madore (1993), under the
assumption of small extinction and that stars with ages of $>$2~Gyr 
dominate 
the RGB. Our analysis suggests that a modest further reduction 
in distance is now required to account for increased extinction and a 
mix of stellar population ages as discussed below. 

The
Pegasus dwarf has also played a role in efforts to obtain distances from
luminous stars. Sandage (1986) used Pegasus as one of his calibrators in
exploring the relationship between stellar and parent galaxy
luminosities. This approach was extended by Rozanski \& Rowan-Robinson
(1994) who found (m-M)$_0 =$27.6. However, the recent level of 
star-forming activity should be a key factor in determining the luminosity
distribution of massive young stars in galaxies.
Because Pegasus has a depressed SFR$_0$, 
it is likely that the most luminous stars method
will over-estimate the distance, as is the case for Pegasus. 

In our HST CMDs for Pegasus there are no unique
features from which an accurate distance can be unambiguously obtained,
such as
a blue HB with a narrow luminosity
distribution.  
The shape of the MS and blue loop somewhat constrain the distance to Pegasus.  
If the distance were (m$-$M)$_0 >$26, then 
the mass function makes blue core HeB stars hard to avoid in larger numbers
at the top of blue plume than are observed. 
If (m$-$M)$_0<$23. then we would expect to
see a change in slope of the MS as radiative cores and convective 
envelopes appear in lower mass MS stars.

The luminosity of the 
RGB tip and the RC further  
constrain the range of possible distances to Pegasus.
However, because young and intermediate age stellar populations are
present, the luminosity of the RGB tip can be
uncertain (Lee, Freedman, \& Madore 1993, Saha et al. 1996).  
RGB tip stars on the second ascent of the
RGB during their AGB 
phase can extend the observed RGB tip, and make the galaxy
appear further away.
The RC luminosity  decreases slowly with age (Lattanzio 1991), which
makes it an inaccurate distance estimator by itself, although it
is useful as a consistency check. Therefore by reducing the distance 
to Pegasus, we increase the age of stars in the RC.
By comparison
with NGC~147, whose distance is  reliably measured by Han et al. 
(1997) from 
the location of the  HB (see Figure~9), 
the range of distances for the Pegasus dwarf consistent with 
the luminosities of 
its RGB tip and RC are: 24.2$\le $(m$-$M)$_0 \le$24.6.

There is no
reasonable reddening that can be combined with distances
of (m$-$M)$_0 \geq$ 24.6 that yields an acceptable match 
to the observed WFPC2 BVI color-color diagram for Pegasus.
Note that in some cases we could obtain a fit for either the
BV or VI CMDs, but the same reddening did not fit both diagrams
simultaneously. Thus reddening and distance are correlated and
{\it observations in three or more filters are
essential in cases where we must derive both a distance and reddening
to a galaxy from a CMD} (see also \S4.3.2).

Therefore we find and adopt a new distance of 
(m$-$M)$_0$=24.4$\pm$0.2 or 
D$=$760$\pm 100$~kpc to Pegasus.  Most of this change 
compared to the larger distance of Aparicio (1994) is due to our 
assumption of nearly 0.3 mag more I-band extinction.
Our revised distance suggests that Pegasus is a member of the M~31 
``family'' within the Local Group (Karachentsev 1996).

\subsection{Metallicity}

For the initial discussion in this paper we set 
the metallicity of the whole stellar population in Pegasus to 
be equal to that of the youngest stars. We are reasonably confident 
in our choice of the Z$=$0.001 models 
to represent the younger stars in Pegasus over higher metallicities 
due to the presence of stars between the MS and RGB. These should be 
mainly stars in core HeB, blue loop evolutionary 
phases, despite the absence of a well-defined ``blue loop'' 
morphology. This is probably a result of small number statistics, 
and also is consistent with the O abundance 
measured for the brightest HII region in Pegasus. 

We have no data to measure the metallicities of the intermediate-age or older 
stars which make up the RGB.
Such measurements require the detection of 
an HB,  extension of our photometry over a longer color 
baseline (e.g., into the infrared), use of narrower-band filters, 
or spectroscopy of a statistically complete sample of RGB stars.
We can only say that most of the RGB stars are unlikely to be 
much lower in metallicity than the $\approx 1/20$ solar value we 
assume for young stars, or the RGB would be both narrow and blue, 
as in the Carina dSph galaxy (Smecker-Hane et al. 1994, Da Costa 1997). 
This is clearly not the case in the Pegasus dwarf.

Our SFH models therefore assume  
that the stellar metallicity levels have remained approximately 
constant at Z=0.001 over most of the life of the Pegasus dwarf.
Additional support for this assumption 
comes from theoretical models of dwarf galaxy evolution 
(e.g. Mori et al. 1997,  Ferrara \& Tolstoy 1998,  
Hensler, Theis, \& Gallagher 1998).  These 
predict a rapid initial 
metal enrichment at the epoch of formation of the
galaxy, followed by a plateau. 
Hence the assumption of zero metallicity evolution over
most of the lifetime of these small systems could be reasonable.
Our model for the Pegasus dwarf is more metal-poor than the 
evolving-metallicity star formation history models presented by 
Aparicio, Gallart, \& Bertelli (1997a).
If there is significant metallicity evolution in Pegasus, 
such that metallicity substantially declines with increasing 
stellar age, then we will
over-estimate the metallicity for the older stars 
seen on RGB and thus under-estimate their ages. 

\section{The Recent Star Formation History}

Having determined all the initial conditions for our models,
we now turn to finding a SFH model that best matches
our data.
Model CMDs which might represent Local Group dwarf irregular galaxies 
when observed with WFPC2 on the {\it HST} have been calculated 
by Aparicio et al. (1996). These are given in terms of $M_I$ versus
V$-$I, and so are directly comparable to our WFPC2 observations.
While the predicted CMDs are similar in terms of their general
appearance to the observations (especially model A which has a
constant SFR from 0.02 to 15 Gyr in the past), there are some
notable differences which indicate we cannot simply adopt one 
of these standard SFH models (such as constant SFR) for the 
Pegasus dwarf. In this and the following section 
we model the detailed properties of 
our observed CMDs to derive the SFH for the Pegasus dwarf.

\subsection{Modeling the Main Sequence}

Normally upper MS stars yield an accurate view of the 
recent SFH, at least in galaxies where such  
stars are plentiful (e.g. Dohm-Palmer 
et al. 1997b) or obvious MS turnoffs are seen. 
However, in the Pegasus dwarf the MS is 
barely populated for M$_V <$-2, and we therefore have a poor statistical 
base for modeling the recent SFH. What 
is clear, as it was to earlier investigators, is the relative paucity 
of stars produced in Pegasus during the past Gyr.

In a constant SFR galaxy the 
observed density of MS stars declines rapidly with increasing 
luminosity due to the combination of the IMF and increasing $t_{MS}$ with
decreasing $m$ (see Scalo 1986, Holtzman et al. 1997, Massey 1997).
The interpretation of a galactic SFH therefore is most straightforward
over the time span $t_{MS}(M(min))$ corresponding 
to the MS lifetimes of the {\it least} luminous observable
MS stars.  However, to find even these older MS stars 
we must take an iterative approach and subtract younger 
stars to reveal less luminous, older MS stars (see 
Dohm-Palmer et al. 1997b). This is necessary because younger 
generations will overlay the older, lower mass 
MS stars.  In Pegasus, we reach to
$M_V \approx$0 on the MS, and thus to stars with ages
of $\approx$0.5~Gyr. 

We used two approaches to fitting the MS, which is well 
isolated from other evolutionary phases by WFPC2 photometry.
From the technique presented by  
Dohm-Palmer et al. (1997b), we
calculated the SFR from the counts in the MS luminosity function.  
We fit the observed luminosity function using the stellar 
evolution models of Bertelli et al. (1994). 
These provide a mass-magnitude
and mass-age relation for the MS at a metallicity of Z=0.001 (see
\S6.4). We use a Salpeter IMF correction, and assume
the distance modulus to be 24.4. The calculations were done with the
dereddened data, and a completeness correction has been applied (see
Sec. 6.3). The results shown in Figure 
13 are rather noisy due to the small number 
of MS stars. The SFR has been approximately constant over the past 100~Myr 
at $\sim$2000~$\msun$~Myr$^{-1}$~kpc$^{-2}$, with a minimum in the 
SFR occurring between 100 and 200~Myr ago.  

We also derived a  
recent SFH from statistical fits to the CMD using the methods 
of Tolstoy (1996) as described in \S6.2 and \S7.4. We adopted simple 
estimates for the recent SFH and iterated these until satisfactory 
agreement was achieved between the models and younger components 
of the CMD.  The resulting models for the MS are in Figure 13 
which are derived for the SFR given by the solid line in Figure 15.
With this technique we attempt to fit all of the younger 
stars and thus do not derive a SFH based solely on the MS.  
While we can see evidence for fluctuations in the 
recent SFR over time, 
we do not find statistically significant SFR variations by more than 
about a factor of two.

Aparicio, Gallart, \& Bertelli (1997a) suggest that a factor of 3 
spike occurred in the SFR in the Pegasus dwarf about 100~Myr 
in the past. This would show up in our data as an excess of 
MS stars near M$_I = -$2; this is the location where the blue 
MS becomes ill-defined in our data (see upper panel of Figure 9). 
Our results in Figure 13 agree with those of Aparicio et al. in 
showing a local peak in the SFR at about 100~Myr, but disagree on 
the specifics of the evolution out to about 200 Myr, and at all 
points we must contend with significant amounts of statistical 
noise.  We therefore cannot confirm the Aparicio et al. 100~Myr 
peak in the SFR from measurements of MS stars in our WFPC2 images.

\subsection{Blue Core Helium-Burning Stars}

A second approach to measuring SFH using stars core 
HeB evolutionary phases was pioneered by Payne-Gaposchkin 
(1974), who took advantage of the Cepheid variable 
pulsation period-age relationship to trace the recent history 
of star formation in the Large Magellanic Cloud. 
Our group has also utilized blue core helium burning stars 
to derive SFHs from the numbers of such stars as a function of their 
luminosity (Dohm-Palmer et al. 1997b). 

Stars in their `blue loop'  core HeB evolutionary phases 
are tracers of the recent SFH because their luminosities 
scale as a power of the initial mass, and thus luminosity 
correlates with stellar age. Unfortunately,  
since the recent SFR in the 
Pegasus dwarf is very low and the blue core HeB evolutionary phase
is short-lived compared to the MS, there are only a scattering of
stars in the expected location of blue core HeB stars. 

We used two methods to model the observations of 
blue core HeB stars in the Pegasus dwarf. Tolstoy (1996) 
presented a statistical approach to deriving a recent SFH from 
the properties of short-lived stars on a CMD. Since intermediate 
and high mass stars move rapidly in a CMD during their post-MS 
evolutionary phases, she used star number counts in relatively 
large boxes to as a way to statistically compare model predictions 
with the observations. This approach was applied to our Pegasus 
CMDs with fits being made simultaneously to the MS, core HeB, 
and RGB components of the CMD. We then find that the SFR in 
Pegasus has increased by about a factor of 2 during the 
past $\sim$0.5~Gyr (see Figure~15).
 
Following Dohm-Palmer et al. (1997b) we also selected
candidate blue core HeB stars from our data 
and generated a luminosity function from which we
calculated the SFR. While this approach of focusing on only one 
evolutionary phase within a complex stellar population can yield 
detailed measures of the recent SFH, 
the low number of blue core HeB stars in Pegasus adds
uncertainty to the results. We follow the prescription
of Dohm-Palmer et al. (1997b), using the blue HeB mass-magnitude,
mass-age, and lifetime relations from Bertelli et al. (1994). Again,
we adopt a Salpeter IMF, deredden the data, and apply a completeness
correction to the luminosity function. 

The results are presented in Figure~14. These are plotted in terms of the 
SFR per kpc$^{2}$ averaged over the WFPC2 field of view. Error bars 
reflect only the statistical uncertainties and do not include 
any corrections for intermediate color stars that are not in the 
core HeB phase or are non-members of the Pegasus dwarf.  The small 
numbers of these intermediate color stars in the WF3 field 
(see Figures~3 and 10a) indicates that the level of contaminating 
objects must be low, certainly much less than 50\% of the total sample. 
Therefore while the SFR derived in Figure~14 is an upper limit, it 
should not substantially exaggerate the actual SFR.

The aproximate agreement between the SFRs found from the 
MS and blue core HeB 
stars over the last 100~Myr (2000 and 1350 $\msun$~Myr$^{-1}$~kpc$^{-2}$
respectively; see Figure 13) supports our original assumption that most of the 
luminous stars in the Pegasus dwarf located between the 
MS and RGB are on blue loops. 
This consistency also indicates that rising recent SFRs predicted by 
our models are likely to be a proper description, and is also roughly 
in agreement with the conclusions by Aparicio, et al. (1997a).

Because of the low SFR and short lifetime of the blue HeB phase, we
have no information from this population for time scales 
of less than 60 Myr; this is set for the time for intermediate 
mass stars to evolve from the MS to the core HeB phase.  
However, the model  in Figure~13 indicates that between
0.1 and 0.4~Gyr ago the SFR  was 
roughly constant at a slightly lower level than the recent 
SFR. From 0.4-0.8~Gyr before the present the 
SFR was very close to its current levels.  A similar result is 
derived from the statistical fits to the observed CMDs following 
the Tolstoy-Saha methodology,  as 
shown by the solid line in Figure~15.

Beyond 800 Myr the photometric errors blend the blue core HeB stars with
the RGB, and the star formation history can no longer be determined 
from the properties of stars that are not associated with the RGB. 
However, we 
can still find a range of feasible SFHs by statistically fitting 
the RGB and also requiring that we produce the correct numbers of 
extended-AGB and RC stars.

\section{Long Term Star Formation History}

\subsection{The Red Giant Branch}

A major division in post-MS evolution 
occurs for stars with lifetimes 
of about 1 Gyr and initial masses $\lesssim$2 $\msun$. These  low 
mass stars have degenerate helium cores during their RGB 
evolution and ignite helium burning in a helium flash. This leads to 
the nearly constant luminosity of the tip of the {\it old} RGB,  
which can be a useful distance indicator 
(e.g., Lee, Freedman, \& Madore 1993). 
Following the RGB phase low mass stars form a RC or HB 
during core helium-burning and thereafter increase 
in luminosity along the AGB. 
More massive stars can evolve on the AGB beyond the RGB tip luminosity  
and produce an extended-AGB before ceasing nuclear burning 
(cf. Vassiliadis \& Wood 1993).

As a result of this RGB transition between non-degenerate and degenerate 
ignition of He on the RGB, the luminosity at the tip of 
the RGB depends upon the mix of  
stellar ages. During the transition to a low mass RGB with helium 
flash at the RGB tip, the luminosity of the tip will increase with 
{\it decreasing} stellar mass (e.g., Sweigart, Greggio, \& Renzini 1990). 
Thus for a single age stellar population, the location of the RGB tip 
is predicted to move towards higher luminosity and redder colors as 
a simple stellar population ages between about 0.5 and 1.2 Gyr.
{\it Therefore any galaxy which experienced extensive 
amounts of star formation during the past $\approx$ 1 Gyr 
may have a complex RGB tip.}

The RGB in Pegasus displays 
a moderate range in color; the change in V$-$I color between the location 
of the RC and its roughly-defined tip is only 
0.5--0.6 magnitudes. Such steeply rising  RGBs are characteristic of 
either moderately metal-poor old stars (e.g., see Lee et al. 1993; 
their calibration suggests [Fe/H]$\approx$-1.5) or of  
relatively massive young stars, whose RGB tracks at a given metallicity are 
bluer at than those of low mass stars. 
That Pegasus cannot contain a pure, old 
RGB can be seen from the presence of intermediate mass MS
stars, the intermediate age RC, and the lack of a sharply  
defined RGB tip; old RGB stars may be present in the Pegasus 
dwarf, but will be challenging to cleanly detect. 
  
The difficulty in distinguishing older RGB stars in a
CMD can be illustrated by considering 
ages of stars on the upper part of the RGB for a 
simple constant SFR model. Stellar evolution models
show that the RGB lifetime at luminosities above the RC are almost
independent of initial mass for $M \le 1.5 \Msun$, or for stellar
nuclear burning lifetimes to the RGB tip of more than 1.5 Gyr.
In this case the number of stars on the upper RGB is given by
$$ N(uRGB) \approx
\psi_0 (t_{gal} - 1.5~Gyr) \int_{M_{RT}(min)}^{M_{RT}(max)}
 \phi(m) dm,$$
where $M_{RT}(min)$ is defined as the initial 
mass of stars with lifetimes at the 
RGB tip of the age of the galaxy $t_{gal}$ and 
$M_{RT}(max) \approx 2 \Msun$ are the most massive
stars to reach the RGB tip.  Therefore $N(uRGB)$ 
approaches being proportional
to the age of the galaxy in long-lived systems, and stars born, for
example, in the first 4 Gyr of a 14 Gyr age, constant SFR galaxy will make up
about 25\% of $N(uRGB)$.  Unless these oldest stars have an obvious
signature that cannot be produced by a younger stellar population
component, such as an associated blue HB, they are not 
readily identified as unique contributors to an observed CMD.

The detectability of the oldest stars also depends on the SFH and the
age-metallicity relationship. If the SFR was higher when a galaxy was young,
then there are more old stars, and detection is more likely. Conversely,
galaxies that had increasing SFRs during their youths can be difficult 
places in which to find older stars. For this reason we have not chosen to 
consider more complicated SFH models in which age and metallicity 
both vary. While such models can be constructed 
(e.g., Aparicio, Gallart, \& Bertelli 1997a) and are required in some cases 
(e.g. Einsel et al. 1995, Han, Hoessel, \& Gallagher 1998), 
they are not justified in the Pegasus dwarf where we have yet to 
find any discernible signs of an age-metallicity relationship.
If we were to adopt an age-metallicity correlation, then this ad hoc 
assumption would almost entirely determine our derived SFH for the 
intermediate age and old stellar populations. Measurements of the 
stellar age-metallicity relationship are required to overcome this 
important ambiguity.
 
\subsection{The Red Clump and Related Issues}

Our WFPC2 data show an excess density of stars on the VI CMD
overlaying the RGB near I$=$24.5 (see Figure~6).  These stars are
in core helium-burning evolutionary phases. The distribution of red,  
helium-burning stars depends sensitively on initial mass, 
amount of mass loss,  and metallicity,
but such stars may be classified into three main categories:
intermediate mass stars at the base of the blue loop, 
RC stars which have experienced a helium flash at the tip of the RGB,
or old, red HB stars. The latter possibility seems
unlikely as the mean color of the RC region 
is not offset to bluer colors, as is expected for a true red horizontal
branch (cf. HB seen in NGC147 data, shown in Figure~9).  
Further consideration is given to the two possibilities
that either most of these stars are associated with the RC or are more massive
stars which did not experience helium ignition
in a degenerate core.

Each of these models implies a different SFH for the Pegasus dwarf.
One model produces most of the RC helium-burning stars
by invoking an epoch of enhanced star formation
about 1 Gyr ago (dashed line in Figure 15). 
Stars made at this time have very short red giant
lifetimes, and relatively long core-helium-burning lifetimes in the
red. This model then yields a high density of red stars at the
required luminosity.  It also implies a strong MS turnoff
associated with a relatively high amplitude burst
should exist at about 1 magnitude fainter than our data, and a 
well-populated extended-AGB. This model requires considerable 
fine-tuning as the transition between the RC and blue loop occurs 
over a very small range of mass and is very sensitive to metallicity, 
and we do not see many blue loop stars, which should be plentiful 
in what is essentially a post-burst galaxy. 

The second and preferred interpretation assumes that the observed 
excess of stars superimposed on the lower RGB are a
classical post-RGB RC.  In this case the RC phase lifetimes
are about 85-70 Myr for initial mass 
0.9--2$\Msun$ stars with Z=0.001 (e.g., Seidel, Demarque, \& Weinberg 1987).
The lifetime on the RGB for stars at luminosities
equal to or higher than the RC are
constant at about 50 Myr for star with initial masses of about 1.5 $\Msun$
or less, and decrease as the upper mass bound for
stars that undergo helium flashes at the RGB tip 
is reached (e.g., Sweigart, Greggio, \&
Renzini 1990).  For example, in the Geneva models with Z=0.001 the
RGB lifetime for log(L)$\geq$1.8 is 52-53 Myr for 0.9-1.5 $\Msun$ stars
and drops to 31 Myr for an initial mass of 1.7 $\Msun$. Thus a high
ratio of RC to RGB stars would be a 
strong signature of enhanced star formation within the
past $>$2 Gyr.

Our WFPC2 photometry of stars in the Pegasus dwarf does 
not yield a straightforward determination of the
ratio $N(RGB)/N(RC)$ because the 
observational errors (and thus also incompleteness)
are a significant factor 
at the luminosity of the RC.  However, using models of the upper RGB,  
which include corrections for incompleteness, to
estimate how many RGB stars are superimposed on the RC region, we
estimate $N(RGB) \approx$6000 stars, $N(RC) \approx$9500 stars, or
$N(RGB)/N(RC) \approx$0.6. The lifetimes of these two evolutionary phases 
for stars with ages of $>$2~Gyr predict a
ratio of 0.60.  This agreement implies that {\it most} of the stars
that we now see in the RGB and RC of the Pegasus dwarf were formed
more than $\approx$2 Gyr ago.

\subsection{The Extended Asymptotic Giant Branch}

The Pegasus dwarf resembles other nearby irregular galaxies, including the 
Magellanic Clouds (e.g. Mould \& Reid 1987),  
in having a well-defined extended-AGB (Aparicio \& Gallart 1995).  Stars 
on the extended-AGB are more luminous than the RGB tip, and often have lower 
effective temperatures and therefore redder colors than the RGB tip region. 
M. Aaronson and his collaborators (e.g. Cook, Aaronson, \& Norris 
1986) were among 
the first to demonstrate that luminous AGB stars are useful indicators for the 
presence of intermediate age stars in galaxies beyond the Milky 
Way's family. 

More recently Gallart et al. (1994, 1996) incorporated 
extended-AGB stars in their investigation of the SFH of the Local Group IBm 
galaxy NGC 6822. They emphasized the connection between the detection of 
bright AGB stars and existence of a stellar population with ages between 1 
and 10 Gyr, but also noted some of the uncertainties involved with the 
interpretation of these types of stars. 
In particular the extended-AGB is sensitive to  
stellar metallicities and mass loss rates, neither of which 
can be reliably determined from BVI CMDs of a mixed-age stellar population
where the SFH is unknown.
In Pegasus there are about 100  
stars on the extended-AGB,  yielding a ratio to the number 
of RC stars of about 10$^{-3}$. This is consistent with the theoretical 
predictions for lifetimes of extended-AGB stars produced by 
intermediate age stellar populations, 
which are only $\approx$0.1\% of the helium-burning 
lifetime. 

Unfortunately, the true situation regarding the properties of the extended-AGB may 
be even more complicated.  Guarnieri, Renzini, \& Ortolani (1997) reviewed 
a variety of additional factors that may influence the populations of AGB 
stars in galaxies, including the importance of binary systems. Also, dust 
in the winds of luminous AGB stars affects 
their optical magnitudes and colors; 
the CMD of luminous extended-AGB stars is therefore 
difficult to quantitatively interpret.

\subsection{Summary of the Long Term Evolution}

Pegasus contains a pronounced RGB with a well-populated RC. The tip 
of the RGB is diffuse, as is typical in intermediate-age stellar 
systems. The ratios $N(RC)/N(RGB)$ and $N(extended-AGB)/N(RC)$ are consistent 
with predictions for stellar populations which formed their stars between 
2-6 Gyr before the present.
We conclude that in Pegasus the structure of the 
`red plume' tells us that the bulk of stars now on the RGB 
formed more than 1-2~Gyr ago. In addition, the extended-AGB is too 
sparsely populated to agree with models in which most of the 
star formation occurred took place more recently than 1-2~Gyr 
before the present.  While an older, very metal-poor stellar 
population cannot be excluded, it should comprise a minority of the 
RGB stars seen in our CMD for the Pegasus dwarf. 

Constant metallicity 
models are illustrated in Figure~15. These produce a relatively 
young age for the bulk of the stars, most of which 
form in these models from 2-4 Gyr 
ago, when the average SFR was about ten times its current value.
Incorporating declining metallicity levels in older stars 
may be a more realistic assumption. In this class of model the structure of 
the RGB depends on both age and metallicity. 
A general feature of such models will be a more 
prolonged epoch of star formation.  In our best estimate for this type 
of model, the primary star-forming epoch would extend back to about 
8~Gyr ago. The average SFR then is reduced by about a factor of 
three relative to the constant stellar metallicity model. 
This reflects the longer time span 
of active star formation, and that older stars on the RGB are 
more numerous due to the form of the initial mass function.

A purely bimodal model for the stellar age distributions in which a 
trace $\le$1 Gyr age stellar component is seen in combination with a very old, 
low metallicity stellar population is excluded.  This SFH predicts a 
narrow, relatively blue RGB with a sharply-defined RGB tip, that we 
do not see. A bimodal SFH model also fails to produce the observed 
populous RC that is approximately centered on top of the RGB.

\subsection{Pegasus at Its Peak}

Our `young' model where most of the stars in Pegasus are 
produced during about 2 Gyr reaches its maximum luminosity 
at the end of the main star-forming epoch, about 2 Gyr 
before the present.  
Estimating a current stellar mass for the Pegasus dwarf of 
1-2$\times$10$^7\msun$ from its integrated colors 
and optical luminosity (see Table 1), 
we can then calculate the peak luminosity. 
Stellar population evolution models, such as 
those of M\"oller, Fritze-v. Alvensleben, \& Fricke (1997),  
predict that after 2 Gyr, with a constant SFR, $M/L_V \approx$1. 
At maximum light our rapid astration model for Pegasus would have 
had M$_V \approx -$14. The color depends on 
the metallicity distribution of the stars, but would fall in 
the range of 0.1$\leq$B$-$V$\leq$0.3.  Even at its best Pegasus 
might not have been a notable galaxy.

The SFH models that we have obtained for Pegasus allow  
the primary star-forming phase to extend for up to 8 Gyr.
In this case the galaxy would be about 1 magnitude fainter 2 Gyr 
ago, and its color would be considerably redder, e.g. 
B$-$V$\approx$0.4. The only way for the Pegasus dwarf to have 
been more luminous than a dwarf is for it to 
have experienced a major starburst where a significant fraction 
of the total stellar mass forms in $\approx$0.1~Gyr (Babul \& 
Ferguson 1996).  We have yet to find any 
compelling direct evidence for such an event 
(e.g., remnant populous star clusters
or well-defined MS turn-offs), but we also cannot exclude the 
possibility of a major starburst, especially if it occurred 
more than 2~Gyr ago. However, 
the simplest model for the SFH of the Pegasus dwarf galaxy  
assumes that stars  
formed at relatively constant rate over several Gyr, with the 
main episode of star formation ending about 2~Gyr before the present.

\subsection{Comparisons with Dwarf Spheroidal Galaxies}

The larger of the Local group dSph systems are similar to Pegasus 
in terms of size and luminosity. The Galactic dSphs are well known to 
contain examples of purely old, metal-poor galaxies, such as 
the Draco, Sculptor, and Ursa Minor galaxies (see review by Gallagher 
\& Wyse 1994). However, more recently evidence has been growing for 
recent star formation in dSphs, including the galaxies Carina (Smecker-Hane 
et al. 1994) and Leo I (Lee et al. 1993b, Demers et al. 1994; 
see also Da Costa 1997).  These results prove that some  Galactic companion 
dSph have undergone episodic star formation, whose impact on the 
metallicities of the stellar populations is unclear. The effects of 
the star formation episodes are directly seen in the multiple 
MS turnoffs in Carina, but also leave their mark in 
the form of composite HB/RC structures in the observed CMDs.

Additional examples of complicated SFH in very small galaxies can 
be found in dSph and related objects elsewhere in the Local 
Group.  These galaxies also show a range in SFHs. And I 
(van den Bergh 1972, 1974) likely 
has an intermediate-to-old stellar population, and Da Costa et al. 
(1996) estimate a mean stellar age of about 10 Gyr from their WFPC2 
photometry. On the other hand, the transition Local Group dwarf LGS3 
contains HI gas and probably some intermediate ($<$1~Gyr) 
age stars, as well as more extensive old populations (
Aparicio, Gallart, \& Bertelli 1997b, Mould 1997, 
Young \& Lo 1997).
Evidently whatever physical ``clock'' sets the time scale for star formation 
in these small galaxies is quite variable, and does not simply depend 
on galaxy size or its eventual location within the Local Group 
(e.g., Gallagher \& Wyse 1994, Da Costa 1997).

\section{Discussion and Conclusions}

The Pegasus dwarf irregular galaxy is a nearby, low SFR,   
low mass, dwarf galaxy.  Our revised distance to the Pegasus 
dwarf of 760 kpc places this galaxy in the M31 ``family'', but at 
a projected distance on the sky of 400 kpc, it is unlikely to 
be bound to M31.  Pegasus is probably a small,  
independent member of the Local Group.

The established interstellar gas 
reserves in Pegasus from the measured HI content suggest that the 
gas (including cosmic He)-to-stellar mass ratio is 
M$_g /$M$_* =$~0.2.  Pegasus lacks the raw material to substantially 
enlarge its stellar mass, but is not yet a stellar fossil. 
At the present time its SFR is low, 
SFR$_0 \approx$3$\times$10$^{-4} \msun~$yr$^{-1}$, and the time scale 
to exhaust the gas at this rate is 13 Gyr.  The lifetime 
average SFR is several times higher, so it is 
possible that a return to normal SFRs could result in 
gas exhaustion within a few Gyr.  Pegasus is like many other 
irregular galaxies where star formation is ``down but not 
out'' (see Hunter 1997). 

Our observed CMDs for the Pegasus dwarf obtained with WFPC2 in 
the equivalent of BVI colors show a sparsely populated MS
and a complicated ``red plume'' consisting of a fairly broad 
RGB, a pronounced RC (cleanly detected here for the first time), 
and an extended-AGB. The 
extended-AGB covers much of the optical  
galaxy on ground-based images taken with the WIYN Telescope. 
Our quantitative analysis of these CMDs establishes that: (1) 
The reddening of E(B$-$V)$=$0.15 is higher than previously thought. 
This mainly accounts for our reduction in distance 
from $D=$960 kpc found by Aparicio (1994) to D$=$760$\pm$100~kpc. 
(2) Star formation is ongoing at a low level and has produced 
at least one HII region, a blue MS,  and core HeB blue loop stars. 
(3) The RGB is prominent, densely populated, and relatively blue. 
(4) The presence of a pronounced RC, the width of the RGB, and 
the lack of a well-defined tip of the RGB are all indicators of a 
strong intermediate age stellar population component. 

We have quantitatively modeled the SFH for Pegasus 
by fitting our observed CMDs following the methodologies  
of Tolstoy \& Saha (1996) and Dohm-Palmer et al. 
(1997b). The short term trend in SFR shows a modest decline 
in SFR during the past 0.5~Gyr from the densities of MS and 
blue core HeB stars as functions of luminosity. 
The Pegasus dwarf therefore shows a slowly varying recent 
mean SFR, but nothing that would qualify 
as a recent star burst. 

Our data are not sufficiently deep to yield a unique description 
of the SFH over more than $\approx$1~Gyr. Instead we find a spectrum 
of possible models for the long term SFH of the Pegasus dwarf. 
At one extreme is a rapid evolution model that 
makes most of the stars between 2 and 4 Gyr before the present 
and assumes constant metallicity at Z$=$0.001. At the other 
extreme are 
less well-defined models where metallicity declines in older 
stars; these models have lower SFRs spread over longer time 
intervals, extending perhaps to 10 Gyr in the past. 

The historical optical characteristics of Pegasus depend on its 
SFH. However, for models where the SFR has been roughly constant 
at intermediate ages of a few Gyr, it would have been most luminous
(M$_V \approx-$14) and relatively blue when its SFR began its recent  
decline about 2 Gyr ago.  Even at its peak Pegasus was most 
likely a ``dwarf'' in terms of absolute magnitude.  

A galaxy like Pegasus in our most actively star-forming 
model would be brightest at a lookback time of $\approx$2~Gyr, or a 
redshift of $z \approx$0.2. At this distance a Pegasus-like 
actively star-forming dwarf would have V$\sim$26, blue colors, 
and an angular size of $<$0.5 arcsec. Thus this type of object 
could resemble some of the low redshift, faint blue galaxies 
(e.g. Odewahn et al. 1996).

If the Pegasus dwarf were to lose its 
gas supply, it might structurally resemble a spheroidal dwarf galaxy.
Lo et al. (1993) show the HI velocity field in Pegasus to be fairly chaotic.
They place an upper limit of 5 km~s$^{-1}$ on gas rotation; measure
a local velocity dispersion of about 5 km~s$^{-1}$; and a global
dispersion of 9 km~s$^{-1}$. Interestingly, stars formed from gas with
these random velocities would likely retain a velocity dispersion similar
to that found in the most luminous of the Galactic dSph galaxies 
(e.g., Gallagher \& Wyse 1994). Furthermore, the global rotation 
is sufficiently slow that the ratio of the stellar velocity dispersion to 
rotation velocity  also is likely to be greater than 1, and so the 
older stellar body of the Pegasus dwarf probably has a spheroidal rather 
than thin disk structure, and resembles  
a low luminosity dE or dwarf spheroidal galaxy. Measurements of stellar 
kinematics would test this possibility, and 
could be done through observations of the  
relatively bright extended-AGB stars in the Pegasus dwarf.

Stellar photometry 
at the level of the HB will similarly enhance our 
ability to constrain the long term SFH of the Pegasus dwarf.
The blue HB that is seen in 
And I (Da Costa et al. 1996) and NGC~147  
(Han et al. 1997) demonstrate the existence 
of old, metal-poor stellar populations, while also 
yielding reliable distances. Evidently these two spheroidal 
galaxies made most of their stars over a time span of several 
Gyr centered at about 8-10~Gyr before the present.
In comparing the Pegasus dwarf to these galaxies we need  
to establish the importance of its 
very old ($>$8~Gyr), metal-poor stellar population component, 
which define the level of star forming activity during the 
primary epoch of giant galaxy formation. By way of contrast,  
in the LMC such stars are rare (Olszewski,  Suntzeff, \& Mateo 1996), 
and we would like to know if this is a general distinction 
between the irregular and spheroidal dwarf galaxies.  This 
information is essential in charting possible relationships 
between present-epoch dwarf galaxies, and faint blue galaxies 
seen at moderate-to-high redshifts corresponding to lookback 
times exceeding several Gyr.

The improvements in the quantitative analysis of  
stellar populations in nearby galaxies made possible by WFPC2 on the 
{\it HST} prove that despite its small mass, 
the Pegasus dwarf has had an extended 
evolutionary history; this galaxy has likely supported star 
formation for at least 3~Gyr. 
The attractive idea that very small galaxies should 
experience a single primary and possibly cataclysmic episode of star 
formation during their lifetimes does not appear to hold in 
the Pegasus dwarf or many other low mass Local Group galaxies. 
We are then left with the intriguing problems of understanding 
what physical mechanisms allow some small galaxies to produce 
their stars slowly and without violence, and how this relates 
to the variety of stellar age mixes and gas contents found in extreme 
dwarf galaxies in the nearby universe.

\bigskip

Support for this work was provided by NASA through 
grant GO-5915 from the Space Telescope Science Institute, which is 
operated by AURA, Inc., under NASA contract NAS 5-26555. 
This research has made use of the NASA/IPAC Extragalactic Database (NED) 
which is operated by the
Jet Propulsion Laboratory, California Institute of Technology, 
under contract with the National
Aeronautics and Space Administration. 
ET thanks Ralph Bohlin for research support during the early phases 
of this project.  JSG expresses his appreciation for partial travel support 
from ST-ECF which made visits to Garching possible. We thank an 
anonymous referee for comments which improved this paper.

\newpage

\begin{deluxetable}{lcc}
\tablenum{1}
\tablecaption{Properties of the Pegasus Dwarf Irregular Galaxy}
\tablehead{
\colhead{Property} &
\colhead{Value} &
\colhead{Reference}}
\startdata
E(B$-$V)        &        0.15   &       This Paper      \nl
B$_T^0$         &       12.61   &       NED             \nl
(B$-$V)$_0$     &       $\approx$0.47 & NED             \nl
D               &       760~kpc $\pm$100 kpc &  This Paper \nl
M$_B$           &       -11.8   &       This Paper      \nl
M(HI)           &       3.1$\times$10$^6 \msun$ & Hoffman et al. 1996 \nl
M(HI)/L$_B$     &       0.4    &       \nodata            \nl
A$_0$           &       1.7~kpc&       Holmberg 1958       \nl
\enddata
\end{deluxetable}

\begin{figure}
\figurenum{1}
\caption{Mosaic of the WFPC2 F555W (`V') image of the central 
regions of the Pegasus dwarf irregular galaxy.  
This galaxy is highly resolved by WFPC2, and crowding is not a major 
factor in the photometric accuracy.  Note that several background galaxies 
are visible; evidently Pegasus is obscuring a relatively rich 
group of more distant galaxies.}
\end{figure}

\begin{figure}
\figurenum{2}
\caption{Section of the 
WFPC2 F555W image of the Pegasus dwarf galaxy is shown from 
the WF4 camera. This detail of the data illustrates the low stellar 
densities at our magnitude limit and the high visibility of background 
galaxies seen through the Pegasus dwarf.} 
\end{figure}

\begin{figure}
\figurenum{3}
\caption{Observed color-magnitude diagrams for the Pegasus 
dwarf galaxy are presented for each of the four 
WFPC2 CCDs in terms of V magnitude and B$-$V color}
\end{figure}

\begin{figure}
\figurenum{4}
\caption{Same as Figure 3 but for I and V$-$I.}
\end{figure}

\begin{figure}
\epsscale{0.8}
\figurenum{5-left}
\caption{The combined color-magnitude diagrams derived from the WFPC2 data 
sets are shown. These data are the basis of our analysis of the 
SFH of the Pegasus dwarf galaxy.}
\end{figure}

\begin{figure}
\figurenum{5-right}
\caption{The combined color-magnitude diagrams derived from the WFPC2 data 
sets are shown. These data are the basis of our analysis of the 
SFH of the Pegasus dwarf galaxy.}
\end{figure}

\begin{figure}
\figurenum{6}
\caption{These plots illustrate the photometric errors for each of the 
three bands for the Wide Field CCDs which contain most of the measured 
stars.}
\end{figure}

\begin{figure}
\figurenum{7}
\caption{Pegasus observed in the V-band with the WIYN 3.5-m telescope 
shows the bright core where star formation is occuring and the more 
extended, dE-like main body of the galaxy. 
Extended  AGB stars and the more luminous RGB members are well-resolved 
over much of the galaxy in this 0.6-arcsec seeing image.  Our field of 
view is 6.7 arcmin on a side, corresponding to 1.5~kpc for a distance of 
760~kpc. The location of the WFPC2 observations is also shown.}
\end{figure}

\begin{figure}
\figurenum{8}
\caption{Color-magnitude diagram in the instrumental i and v$-$i colors 
from photometry of stars in our WIYN images. These data have been approximately 
transformed to standard magnitudes using our WFPC2 results.}
\end{figure}

\begin{figure}
\figurenum{9}
\caption{Pegasus I,V$-$I CMD is the top plot, and for comparison
the NGC~147 outer field WFPC2  
I,V$-$I CMD from Han et al. (1997) is shown below it.}
\end{figure}

\begin{figure}
\figurenum{10a}
\caption{These plots show the projected densities of stars selected 
from different regions of our WFPC2 CMDs. This diagram gives 
densities of younger stars on the main sequence and in core helium-burning 
evolutionary stages within an 8-arcsec smoothing region.} 
\end{figure}

\begin{figure}
\figurenum{10b}
\caption{Same as for Figure 10a but showing the locations of RGB stars, 
which are less centrally-concentrated than the younger stars.}
\end{figure}

\begin{figure}
\figurenum{10c}
\caption{This density plot for extended AGB stars is derived 
from our WIYN data and has the same 
orientation as Figure 7. It was made with the same smoothing scale used 
for the WFPC2 density plots, and 
shows a relatively clumpy distribution of AGB stars in the inner galaxy.
Poor statistics produced low amplitude features in the outer parts 
of the galaxy.  Note that the star-forming region is offset with respect 
to the distribution of AGB stars.}
\end{figure}

\begin{figure}
\figurenum{11}
\caption{A comparison is shown between Geneva and Padova stellar 
evolution tracks.  The Padova models were interpolated to the same 
equivalent evolutionary points as were used to make the Geneva 
tracks. Note the offset between the RGB tip locations for these 
two sets of models.}
\end{figure}

\begin{figure}
\figurenum{12}
\caption{IRAS 100$\mu$m map of a region centered on the Pegasus 
dwarf that is marked by a small box.  The field is 1 degree square 
with 15~arcmin$^2$  
pixels}
\end{figure}

\begin{figure}
\figurenum{13}
\caption{The upper panel shows the SFR versus time derived with 
the Dohm-Palmer models for the MS and from blue core HeB stars. 
Error bars refer only to statistical uncertainties. The lower 
panel gives cumulative SFRs, and demonstrates the basic 
agreement between the MS and blue core HeB methods.}
\end{figure}

\begin{figure}
\figurenum{14}
\caption{The SFR is shown as calculated from the luminosity 
function for stars identified 
as being in the blue core-HeB phase.   
The bins are 50 Myr wide. The data have been dereddened and
corrected for incompleteness. The assumed IMF has a Salpeter
slope. The low numbers of candidate blue HeB stars implies a low SFR,
and contributes fairly large errors to the calculation. For times 
beyond 800 Myr the photometric
errors blend the blue HeB stars and the RC and RGB, making the 
calculation uncertain.}
\end{figure}

\clearpage

\begin{figure}
\figurenum{15}
\caption{Observed CMD for Pegasus (upper left) is compared with 
two synthetic models computed with the Tolstoy and 
Saha statistical fitting method. 
The lower model is for our best estimate SFH 
fo Z=0.001 and the upper 
right model for Z=0.004 based on the Geneva stellar evolution tracks.
While neither of this pair of models properly fits the blue loop stars, 
the lower metallicity is a statistically better option because it 
includes some stars with intermediate colors which are likely 
candidates for stars on blue loops.}
\end{figure}


\begin{figure}
\figurenum{16}
\caption{SFH models derived from statistical fits to our WFPC2 
photometry of the Pegasus dwarf are shown in terms of {\it 
relative} SFRs. The solid curve is the recent SFH based on 
the MS, blue loop stars, and the blue side of the RGB with Z=0.001. 
This is the 
best constrained component of the model. The short dashes show a 
high metallicity (Z$=$0.004) model in which most stars are 
less than 1~Gyr old. This model fits the shape of the CMD, 
but produces no stars as blue as the observed blue loop 
candidates, and has difficulty explaining the AGB population.
The long dashes is a speculative SFH model designed to 
match the RGB. Since 
we assume constant metallicty the width in color of the RGB 
depends only on the range of stellar ages. If metallicity 
declines with increasing age, then the height of the peak 
declines and its width increases; i.e. the mean SFR drops and 
the galaxy becomes older.}
\end{figure}

\end{document}